\documentclass[nohyper,11pt,letterpaper]{JHEP3}
\usepackage[dvips]{epsfig}
\usepackage{epsf,amsfonts,amssymb}

\newcommand{\eqn}[1]{(\ref{#1})}
\newcommand{\be}{\begin{equation}}
\newcommand{\ee}{\end{equation}}
\newcommand{\ben}{\begin{displaymath}}
\newcommand{\een}{\end{displaymath}}
\newcommand{\bea}{\begin{eqnarray}}
\newcommand{\eea}{\end{eqnarray}}
\newcommand{\bean}{\begin{eqnarray*}}
\newcommand{\eean}{\end{eqnarray*}}
\newcommand{\nn}{\nonumber \\}
\newcommand{\ba}{\begin{array}}
\newcommand{\ea}{\end{array}}
\newcommand{\bi}{\begin{itemize}}
\newcommand{\ei}{\end{itemize}}

\def\l {\lambda}
\def\a {\alpha}
\def\b {\beta}

\def\d {\delta}
\def\s {\sigma}
\def\e {\epsilon}

\renewcommand{\t}{\theta}

\def\be{\begin{equation}}
\def\ee{\end{equation}}

\def\e{\epsilon}

\def\a{\alpha}
\def\b{\beta}

\newcommand{\cale}{\mbox{${\cal E}$}}

\newcommand{\caln}{\mbox{${\cal N}$}}

\newcommand{\calq}{\mbox{${\cal Q}$}}

\newcommand{\bbr}[1]{\mbox{${\mathbb R}^{#1}$}}

\font\mybb=msbm10 at 8pt
\def\bb#1{\hbox{\mybb#1}}
\def\bR {\bb{R}}

\newcommand{\pa}{\partial}
\newcommand{\fc}{\frac}
\newcommand{\w}{\wedge}

\newcommand{\sac}{\, , \qquad}

\newcommand{\ie}{{\it i.e.}}

\newcommand{\ep}{\ensuremath{{\eta'}}}

\newcommand{\cone}{\ensuremath{{C_{\it 1}}}}
\newcommand{\ftwo}{\ensuremath{{F_{\it 2}}}}
\newcommand{\csev}{\ensuremath{{C_{\it 7}}}}
\newcommand{\feig}{\ensuremath{{F_{\it 8}}}}
\newcommand{\osev}{\ensuremath{{\omega_{\it 7}}}}

\newcommand{\tr}{\mbox{Tr}}

\newcommand{\mt}[1]{\textrm{\tiny #1}}
\def\ls{\ell_s}
\def\nc {N_\mt{c}}
\def\nf {N_\mt{f}}

\newcommand{\mq}{\ensuremath{m_\mt{q}}}      
\def\ua {U(1)_\mt{A}}
\def\ukk {U_\mt{KK}}
\newcommand{\mkk}{M_\mt{KK}}
\def\gym {g_\mt{YM}}
\newcommand{\cc}{\langle \bar{\psi} \psi \rangle}
\def\half{{1\over 2}\,}

\newcommand{\ra}{\rightarrow}

\title{\LARGE The Holographic Life of the $\eta'$}

\author{Jos\'e L. F. Barb\'on,$^{a}$ Carlos Hoyos,$^{b}$
  David Mateos,$^{c}$ and Robert C. Myers$\,^{c,d}$ \\
  $^a$ Theory Division, CERN, CH-1211, Geneva 23, Switzerland \\
  $^b$ Instituto de F\'{\i}sica Te\'orica UAM/CSIC,  C-XVI \\
       and Departamento de F\'{\i}sica Te\'orica, C-XI \\
       Universidad Aut\'onoma de Madrid, E-28049--Madrid, Spain \\
  $^c$ Perimeter Institute for Theoretical Physics \\
       Waterloo, Ontario N2J 2W9, Canada \\
  $^d$ Department of Physics, University of Waterloo  \\
       Waterloo, Ontario N2L 3G1, Canada \\

E-mail: \email{barbon@cern.ch, c.hoyos@uam.es,
  dmateos@perimeterinstitute.ca, rmyers@perimeterinstitute.ca}}

\abstract{In the string holographic dual of large-$\nc$ QCD with
$\nf$ flavours of \cite{KMMW03}, the $\eta'$ meson is massless at
infinite $\nc$ and dual to a collective fluctuation of $\nf$
D6-brane probes in a supergravity background. Here we
identify the string diagrams responsible for the generation
of a mass of order $\nf/\nc$, consistent with the
Witten-Veneziano formula, and show that the supergravity limit
of these diagrams corresponds to mixings with pseudoscalar glueballs.
We argue that the  dependence on the
$\theta$-angle in the supergravity description
 occurs only through the combination
$\theta + 2\sqrt{\nf} \, \eta' / f_\pi$, as dictated by the $\ua$
anomaly. We provide a quantitative test by computing the linear term
in the $\eta'$ potential in two independent ways, with perfect
agreement.}


\preprint{}

\begin{document}

{\vskip 1cm}

\section{Introduction}

In QCD with three light flavours of quark,
$m_\mt{u}, m_\mt{d}, m_\mt{s} \ll \Lambda_\mt{QCD}$,
there is a very succesful model
of light meson phenomenology in terms of the spontaneous breaking of
the chiral $SU(3)_\mt{L} \times SU(3)_\mt{R}$  flavour symmetry down
to the diagonal subgroup.  In the same context, the spontaneous
breaking of the axial  $\ua$  group would imply the existence of a
neutral pseudoscalar meson with the quantum numbers of the $\ep$ meson
and mass $m_\ep <\sqrt{3}\, m_\pi$. The measured mass of the $\ep$
meson, close to $1$ GeV, exceeds this bound by a large amount, leading
to the so-called `$U(1)$ problem' \cite{GMOR68, Weinberg75}.

Quantum mechanically, the $\ua$ symmetry is broken by the anomaly,
proportional to $\tr\, F\wedge F$, which in turn means that the $U(1)$
problem is tied to the dependence of physical quantities on the
$\theta$-angle of QCD. In particular, the $\ep$ meson can only be lifted by
non-perturbative effects, since the anomaly itself is a total derivative, and
thus inocuous in perturbation theory.

Because of the anomaly, the effective CP-violating phase is the combination
$\theta + {\rm arg}\,(\,{\rm det}\,\mq \,)$, where $\mq$ denotes the
quark mass matrix for $\nf$ flavours. Hence, normalizing the would-be
$\ua$ Goldstone boson  by the global phase $e^{i\phi}$ of the
$U(\nf)_\mt{A}$ Goldstone-boson matrix $\Sigma$, the anomaly constrains
the low-energy effective potential of the phase field to depend on the
combination  $\theta + \nf \,\phi$ in the chiral limit, $\mq =0$.
For example, a dilute gas of instantons generates a potential of the
form (c.f. \cite{thooft})
\be
\label{potinst}
V(\Sigma)_\mt{inst} = A \,e^{i\theta} \,\det \,\Sigma + {\rm h.c.} \,,
\ee
where $A \sim \exp(-8\pi^2 /\gym^2)$. In the large-$\nc$ limit, this potential
is exponentially supressed. However, it was shown by Witten \cite{Witten79} (see
also \cite{Veneziano79, clasicos}) that a non-trivial $\theta$-dependence
within the $1/\nc$ expansion of the pure Yang--Mills (YM) theory implies
a potential  of the form
\be
\label{scau}
V(\Sigma)_\mt{WV} = {1\over 2} \,\chi_\mt{YM}\,
\left(\theta -i\,\log\,{\rm det}\,\Sigma\, \right)^2
\ee
to first non-trivial order in the $1/\nc$ expansion (generated by a
non-perturbative resummation of OZI-supressed quark annihilation
diagrams \cite{RGG75, Witten79, Veneziano79}).
The constant $\chi_\mt{YM}$ is the topological susceptibility of the
{\it pure} YM theory,
\be
\label{topsus}
\chi_\mt{YM} = {d^2 \,\cale_\mt{vac} \over d\,\theta^2}
\Big |_{\nf =0, \;\theta=0} \,,
\ee
to leading order in the $1/\nc$ expansion. More generally, the large-$\nc$
scaling of the vacuum energy density in the pure YM theory is
\be
\label{pg}
\cale_\mt{vac} = \nc^2 \,F(\theta/\nc) \,,
\ee
where the function $F(y)$ has a Taylor expansion with coefficients of
$O(1)$ in the large-$\nc$ limit, and it  should be multivalued under
$\theta\rightarrow \theta+2\pi$ in order for the $\theta$-angle to be
defined with $2\pi$ periodicity.  Then, applying the substitution
$\theta \rightarrow \theta + \nf \,\phi$ dictated by the anomaly, we
find a potential of the general form
\be
\label{pot}
V(\phi) = \nc^2 \, F\left({\theta + \nf \,\phi \over \nc}\right) \,.
\ee
Notice that the multivalued nature of $\theta$-dependence in the large-$\nc$
limit of pure YM theory is tied to an analogous `multibranched' nature of the
$\ep$ potential, already apparent by the contrast between (\ref{potinst}) and
(\ref{scau}). The $\ep$ mass is obtained by selecting the quadratic term and
introducing the canonically normalized $\ep$ field:\footnote{Note that the present
normalization is consistent with \cite{Witten79}, however, this differs from that
used in \cite{KMMW03}: $f_\pi$\cite{KMMW03}$=f_\pi$\cite{Witten79}$/2$.}
\be
\label{cann}
\phi (x)= {2 \over f_\pi \,\sqrt{\nf}} \,\ep (x) \,,
\ee
where $f_\pi$ is the pion decay constant; since
$f_\pi = f_\ep + O(1/\nc)$, we will not distinguish between the two.
This results in the famous Witten--Veneziano formula
\be
\label{wv}
m_\ep^2 = {4 \nf \over f_\pi^2} \,\chi_\mt{YM} \,.
\ee
Since $f_\pi \sim \sqrt{\nc}$, we get a mass-squared of $O(\nf / \nc)$.

In the same fashion, one can also derive soft-$\ep$ amplitudes by
applying the substitution
$\theta \rightarrow \theta + 2\sqrt{\nf} \,\eta' /f_\pi$ to the
$\theta$-dependence of pure-glueball amplitudes. We can specify
not only the low-energy effective action of the pseudo-Goldstone
field $\ep$,  but also a large-$\nc$ effective Lagrangian featuring
glueballs and mesons with masses of $O(1)$ in the large-$\nc$ limit,
together with a light $\eta'$ meson with mass of $O(1/\nc)$.

In string descriptions of large-$\nc$ gauge theories, such as AdS/CFT
models, it should be possible to verify this scenario by direct
inspection of the low-energy effective action of the
string theory in the AdS-like background, either at the level of the
classical supergravity approximation (glueball-meson spectrum) or at
the level of string loop corrections. In particular, one should find
the potential (\ref{pot}) as part of the effective action in the
background geometry.

As we will review below, the first part of this check was carried out
by Witten \cite{Witten98b}, who studied the $\theta$-dependence of an
AdS-like model \cite{Witten98a} dual to a non-supersymmetric,
confining cousin of pure YM theory. Introducing  $\theta$-dependence
through Ramond--Ramond (RR) fields, Witten derived the analog of
(\ref{pg}) for this model, with the result
\be
\label{tow}
\cale_\mt{vac}^{(k)} = \nc^2 \,F_k(\theta/\nc) =
{1\over 2} \,\chi_g \,(\theta + 2\pi k)^2 + O(1/\nc)
\ee
to leading order in the $1/\nc$ expansion,
where the integer $k$ labels the $k$-th stable `vacuum'.
Minimizing over $k$ for a given value of $\theta$ selects the true
vacuum and restores the $2\pi$ periodicity. The $O(1)$ constant
$\chi_g$ is the topological susceptibility in this model.

In order to complete the check we need a generalization of this setup
that incorporates flavour degrees of freedom in the chiral limit.
In the large-$\nc$ limit it  should also incorporate a {\it massless},
pseudoscalar Goldstone boson that can be identified with the $\eta'$ field.
Following the general ideas of \cite{flavour}, a model with exactly
these properties was constructed in \cite{KMMW03} by introducing
flavour degrees of freedom via D6-brane probes embedded in the
previous background.\footnote{Following the ideas of \cite{flavour},
  meson physics has been studied in the context of AdS/CFT in
  \cite{mesons}.} In this note we investigate the $\ep$ physics in
this model.

We first argue that the introduction of D6-branes corresponding to
massless quarks allows the dependence of the supergravity
description on the microscopic
$\theta$-angle to be shifted away, precisely as expected on field
theory grounds. We then discuss the kind of string loop corrections
that must be responsible for the generation of the anomaly-induced
potential (\ref{scau}), in a string analog of the old Isgur-de
R\'ujula-Georgi-Glashow mechanism
\cite{RGG75}. Although we are unable to provide an independent
stringy calculation of the $\ep$ mass, we show that, in the
supergravity limit, the leading Wess--Zumino coupling of the
D6-brane probes to the RR background fields induces the right
structure of mixings between the $\ep$ meson and pseudoscalar
glueballs. In section \ref{quanti} we present a non-trivial
quantitative check of this scenario by computing the linear term of
the potential (\ref{scau}) in two independent ways, with precise
agreement.

In order for this paper to be self-contained, we have included, in
section \ref{summary}, a summary of the aspects of
\cite{KMMW03, Witten98b, Witten98a} that are needed in the rest of
the paper. Readers who are familiar with these can go directly to
section \ref{anomaly}.

\section{The Model} \label{summary}

A proposal to realize a holographic dual of four-dimensional,
non-supersymmetric, pure $SU(\nc)$ YM theory was made in \cite{Witten98b}.
One starts with $\nc$ D4-branes in the type IIA Minkowski vacuum
$\bbr{9} \times S^1$. The D4-branes wrap the compact direction, of
radius $M_\mt{KK}^{-1}$, and anti-periodic boundary
conditions are imposed for the worldvolume fermions on this circle.
Before compactification, the D4-brane theory is a five-dimensional,
supersymmetric $SU(\nc)$ gauge theory whose field content
includes fermions and scalars in the  adjoint representation of
$SU(\nc)$, in addition to the gauge fields. At energies much below
the compactification scale, $M_\mt{KK}$, the theory is effectively
four-dimensional. The anti-periodic boundary conditions break all of
the supersymmetries and give a tree-level mass to the fermions, while
the scalars also acquire a mass through one loop-effects. Thus, at
sufficiently low energies, the dynamics is that of four-dimensional,
massless gluons.

If the type IIA vacuum is such that there is a non-trivial holonomy
around the circle for the RR one form, $\cone$, then the Wess-Zumino
coupling on the D4-branes,\footnote{We adopt a nonstandard
convention where the field components $(\cone)_\mu$ have
dimensions of length$^{-1}$, \ie,
$\cone$\cite{dielectric}$=g_s\ell_s\,\cone$[present]. Hence as
forms, $\cone$ and $\ftwo$ are both dimensionless which will
simplify various expressions in the following. Note that with
these conventions, the forms $\csev$ and $\feig$, defined by the
usual duality relation $\feig = *\,\ftwo$ in subsequent sections,
both have dimensions of length$^{6}$.}
\be
\label{thec}
{1\over 8\pi^2} \int_{\bR^4 \times S^1} \cone \wedge
\tr \, F\wedge F \,,
\ee
induces a $\t$-term in the gauge theory with
\be
\label{ett}
\theta = \int_{S^1} \cone \,.
\ee

The D4-brane system above has a dual description in terms of
string theory in the near-horizon region of the associated
(non-supersymmetric) supergravity background.
Using this description, Witten showed \cite{Witten98b}
that the $\t$-dependence of the vacuum energy of the YM theory has
precisely the form expected on field theory grounds, as reviewed in
the Introduction.

In order to explore the new physics associated to the $\eta'$
particle, we need to extend Witten's construction in such a way that,
in the limit in which the KK modes would decouple, the only additional
degrees of freedom would be $\nf$ flavours of fundamental, massless
quarks.\footnote{As usual
in AdS/CFT-like dualities, this limit is not fully realisable within
the supergravity approximation; see \cite{KMMW03} for a more
detailed discussion.} Such an extension was proposed in
\cite{KMMW03}, following the general strategy of adding fundamental
matter to AdS/CFT by adding D-brane probes \cite{flavour}. The
construction is as follows.

Consider adding $\nf$ D6-branes to the original system, oriented as
described by the array
\be
\begin{array}{rccccccccccl}
\nc \,\, \mbox{D4:}\,\,\, & 0 & 1 & 2 & 3 & 4 & \_ & \_ & \_ & \_ & \_ & \, \\
\nf \,\, \mbox{D6:}\,\,\, & 0 & 1 & 2 & 3 & \_ & 5 & 6 & 7 & \_ & \_ & \, .
\ea
\label{intersection}
\ee
The original gauge fields and adjoint matter on the D4-branes
arise from the light modes of the 4-4 open strings, and propagate in
five dimensions. In contrast, the light modes of the 4-6 open strings
give rise to $\nf$ hypermultiplets in the fundamental
representation of $SU(\nc)$ that propagate only along the four
directions common to both branes.\footnote{We emphasize that these
fields are intrinsically four-dimensional, \ie, they do {\it not} propagate
along the circle direction.} Each hypermultiplet consists of one
Dirac fermion, $\psi = \psi_\mt{L} + \psi_\mt{R}$, and two complex
scalars. The addition of the D6-branes leaves $\caln =2$ unbroken
supersymmetry (in four-dimensional language). This ensures that there is
no force between the D4- and the D6-branes, and hence that they can be
separated in the 89-plane. The bare mass of the hypermultiplets,
$\mq$, is proportional to this separation. If the D6-branes lie at the
origin in the 89-plane, then the system enjoys
a $\ua$ symmetry associated to rotations in this plane. A crucial
fact in the construction of \cite{KMMW03} is that,
in the gauge theory, this symmetry acts on the
fundamental fermions as a {\it chiral} symmetry, since it rotates
$\psi_\mt{L}$ and $\psi_\mt{R}$ with opposite phases. Hence the
$\ua$ symmetry acts on the relevant fields as
\be
\label{rsim}
X_8 + iX_9 \ra e^{i\alpha}\,(X_8+iX_9) \sac
\psi_\mt{L} \ra e^{i\alpha/2} \, \psi_\mt{L} \sac
\psi_\mt{R} \ra e^{-i\alpha/2} \, \psi_\mt{R} \,.
\ee

As discussed above, identifying the 4-direction with period
$2\pi/M_\mt{KK}$, and with anti-periodic boundary conditions for
the D4-brane fermions, breaks all of the supersymmetries and
renders the theory effectively four-dimensional at energies $E \ll
M_\mt{KK}$. Further, the adjoint fermions and scalars become
massive. Similarly, we expect loop effects to induce a mass for the
scalars in the fundamental representation. Generation of a mass for
the fundamental fermions is, however, forbidden (in the strict
large-$\nc$ limit) by the existence of the chiral $\ua$ symmetry
above. Therefore, at low energies, we expect to
be left with a four-dimensional $SU(\nc)$ gauge theory coupled to
$\nf$ flavours of fundamental quark.

In the so-called `probe limit', $\nf \ll \nc$, a holographic
description of this theory is obtained by replacing the D4-branes
by their supergravity background. The condition $\nf \ll \nc$ ensures
that the backreaction of the D6-branes on this background is
negligible, and hence that they can be treated as probes.
The D6-brane worldvolume fields (and, more generally, all open string
excitations on the D6-branes) are dual to gauge-invariant field
theory operators constructed with at least two hypermultiplet fields,
that is, meson-like operators; of particular importance here will be
the quark bilinear operator, $\bar{\psi} \psi \equiv \bar\psi_i \psi^i$,
where $i=1, \ldots , \nf$ is the flavour index.

Having reviewed the general construction, we now provide some of the
details from \cite{KMMW03} that will be needed in the following sections.

The supergravity background dual to the $\nc$ D4-branes
takes the form
\begin{eqnarray}
ds^{2} &=& \left(\frac{U}{R}\right)^{3/2} \left( \eta_{\mu \nu} \,
dx^\mu dx^\nu + f(U) d\tau^{2} \right) + \left(
\frac{R}{U}\right)^{3/2} \frac{dU^{2}}{f(U)} +
R^{3/2} U^{1/2} \, d\Omega_{\it 4}^{2} \,, \label{metric} \\
e^{\phi} &=& g_s \left( \frac{U}{R}\right)^{3/4}
\sac F_{\it 4} = \frac{\nc}{\Omega_{\it 4}} \, \varepsilon_{\it 4} \sac
f(U) = 1-\frac{\ukk^{3}}{U^{3}} \,.
\label{metric1}
\end{eqnarray}
The coordinates $x^\mu=\{ x^0, \ldots , x^3\}$ parametrize
$\bbr{4}$, and correspond to the four non-compact directions along
the D4-branes, as in \eqn{intersection}, whereas $\tau$
parametrizes the circular 4-direction on which the branes are
compactified. $d\Omega_{\it 4}^2$ and $\varepsilon_{\it 4}$ are
the $SO(5)$-invariant line element and volume form on a unit
four-sphere, respectively, and $\Omega_{\it 4}=8\pi^2/3$ is its
volume. $U$ has dimensions of length and may be thought of as a
radial coordinate in the 56789-directions transverse to the
D4-branes. Since the $\tau$-circle shrinks to zero size at
$U=\ukk$, to avoid a conical singularity $\tau$ must be identified
with period
\be
\d \tau = \fc{4 \pi}{3} \, \fc{R^{3/2}}{\ukk^{1/2}} \,.
\label{deltatau}
\ee
Under these circumstances the supergravity solution above is regular
everywhere. $U$ and $\tau$ parametrize a `cigar' (as opposed to a
cylinder). That is, the surface parametrized by these coordinates
is topologically a plane.
The solution is specified by the string coupling constant, $g_s$,
the Ramond--Ramond flux quantum (\ie, the number of D4-branes),
$\nc$, and the constant $\ukk$. (The remaining parameter is given by
$R^3 =  \pi g_s \nc\,\ell_s^3$, with $\ell_s$ the string length.)
If $\ukk$ is set to zero, the solution (\ref{metric}, \ref{metric1})
reduces to the extremal,
1/2-supersymmetric D4-brane solution, so we may say that $\ukk$
characterizes the deviation from extremality.
The relation between these parameters and those of the $SU(\nc)$
dual gauge theory, namely, the compactification scale,
$\mkk=2\pi/\d\tau$, and the four-dimensional coupling constant
{\it at} the compactification scale, $\gym$, is \cite{KMMW03}:
\be
R^3 = {1\over2} \fc{\gym^2 \nc \, \ell_s^2}{\mkk} \sac
g_s = {1\over2\pi} \fc{\gym^2}{\mkk \ell_s} \sac
\ukk = {2\over9} \gym^2 \nc \, \mkk \ell_s^2 \,.
\label{inverse}
\ee

In the gravity description, the defining equation \eqn{ett} for the
$\t$-angle must be understood as an asymptotic boundary condition for the RR
one-form at $U \ra \infty$. In other words, we must impose
\be
\label{deft}
\theta + 2\pi k = \lim_{U\to \infty} \int_{S^1} \cone =
\int_{\rm Cigar} \ftwo \,,
\ee
where the $S^1$ is parametrized by $\tau$ and lies at
$U=\mbox{constant}$, as well as at constant positions in $\bbr{4}$ and
$S^4$, and $\ftwo=d\cone$.
Notice that the asymptotic holonomy of $\cone$ is measured over a
contractible cycle of the background geometry. Under these circumstances,
the right-hand side of (\ref{deft}) defines an arbitrary real number,
and we must specify the integer $k$ to respect the angular nature
of $\theta$.

To leading order in $1/\nc$, the solution of the supergravity
equations that obeys the constraint \eqn{deft} is obtained
\cite{Witten98b} simply by adding to \eqn{metric} and
\eqn{metric1} the RR two-form \be \label{wittf} \ftwo = {C \over
U^4} \, (\t + 2\pi k) \,dU\wedge d\tau \,, \ee where $C= 3\,
\ukk^3 /\d \tau$. Inserting this expression into the kinetic
action of the RR forms we get Witten's result for the energy
density
\be
\label{wr}
\cale_\mt{vac}^{(k)}
= {1\over 2 (2\pi)^7\ls^{\,\, 6} V_{\it 4}} \int \ftwo\wedge *
\ftwo = {1\over 2} \,\chi_g \,(\theta+ 2\pi k)^2 \,,
\ee
where $V_{\it 4} = \int d^4 x$. The topological susceptibility is
thus given by (c.f. \cite{HO98})
\be
\label{tosu}
\chi_g = {(\gym^2 \nc)^3  \over   4 \cdot (3\pi)^6} \, \mkk^4 \,.
\ee
The generation of a topological susceptibility of
$O(1)$ constrasts with naive expectations based on an instanton gas
picture. In this model, one can explicitly check that the
semiclassical approximation based on a dilute instanton gas does not
commute with the large-$\nc$ resummation provided by the
supergravity approximation \cite{BP99}.

The study of the embedding of the D6-brane probes is greatly
simplified by working in isotropic coordinates in the
56789-directions. Towards this end, we first define a new
radial coordinate, $\rho$, related to $U$ by
\be
U(\rho) = \left(\rho^{3/2} +
\frac{\ukk^3}{4\rho^{3/2}}\right)^{2/3} \,,
\label{isomer}
\ee
and then five coordinates $\vec{z}=(z^5, \ldots, z^9)$ such that
$\rho = |\vec{z}|$ and $d\vec{z} \cdot d\vec{z} = d\rho^2 + \rho^2
\, d\Omega_{\it 4}^2$. In terms of these coordinates the metric
\eqn{metric} becomes
\be
ds^{2} = \left(\frac{U}{R}\right)^{3/2}
\left( \eta_{\mu \nu} \, dx^\mu dx^\nu + f(U) d\tau^{2} \right) +
K(\rho) \, d\vec{z} \cdot d\vec{z} \,,
\label{isometric1}
\ee
where
\be K(\rho) \equiv \fc{R^{3/2} U^{1/2}}{\rho^2}\,.
\ee
Here $U$ is now thought of as a function of $\rho$. To
exploit the symmetries of the D6-brane embedding, we finally
introduce spherical coordinates $\l, \Omega_{\it 2}$ for the
$z^{5,6,7}$-space and polar coordinates $r, \phi$ for the
$z^{8,9}$-space. The final form of the D4-brane metric is then
\be
ds^{2} = \left(\frac{U}{R}\right)^{3/2} \left( \eta_{\mu \nu} \,
dx^\mu dx^\nu + f(U) d\tau^{2} \right) + K(\rho) \, \left(
d\lambda^2 + \lambda^2 \, d\Omega_{\it 2}^2 + dr^2 + r^2 \,
d\phi^2 \right) \,, \label{isometric}
\ee
where $\rho^2 = \l^2 + r^2$. The $\ua$ symmetry corresponds here to
shifts of the $\phi$ coordinate.

In these coordinates the D6-brane embedding takes a particularly
simple form. We use $x^\mu$, $\l$ and $\Omega_{\it 2}$ (or $\s^a$,
$a=0,\ldots,6$, collectively) as worldvolume coordinates.  The
D6-brane's position in the 89-plane is specified as $r=r(\l)$,
$\phi = \phi_0$, where $\phi_0$ is a constant. Note that $\l$ is the
only variable on which $r$ is allowed to depend, by translational
and rotational symmetry in the 0123- and 567-directions,
respectively. We also set $\tau=\mbox{constant}$, as
corresponds to D6-branes localized in the circle direction.

The function $r(\l)$ is determined by the requirement that the
equations of motion of the D6-brane in the D4-brane background be
satisfied. In the supersymmetric limit, $\ukk=0$, $r(\l)=2\pi\ls^{\,2}\,\mq$ is a
solution for any (constant) quark mass $\mq$, as depicted in figure
\ref{d6-embedding}(a); this reflects the BPS nature of the system.
If the quarks are massive then the D6-brane
embedding is not invariant under rotations in the 89-plane and the
$\ua$ symmetry is explicitly broken. If instead $\mq=0$ then the $\ua$
symmetry is preserved.
\FIGURE{\epsfig{file=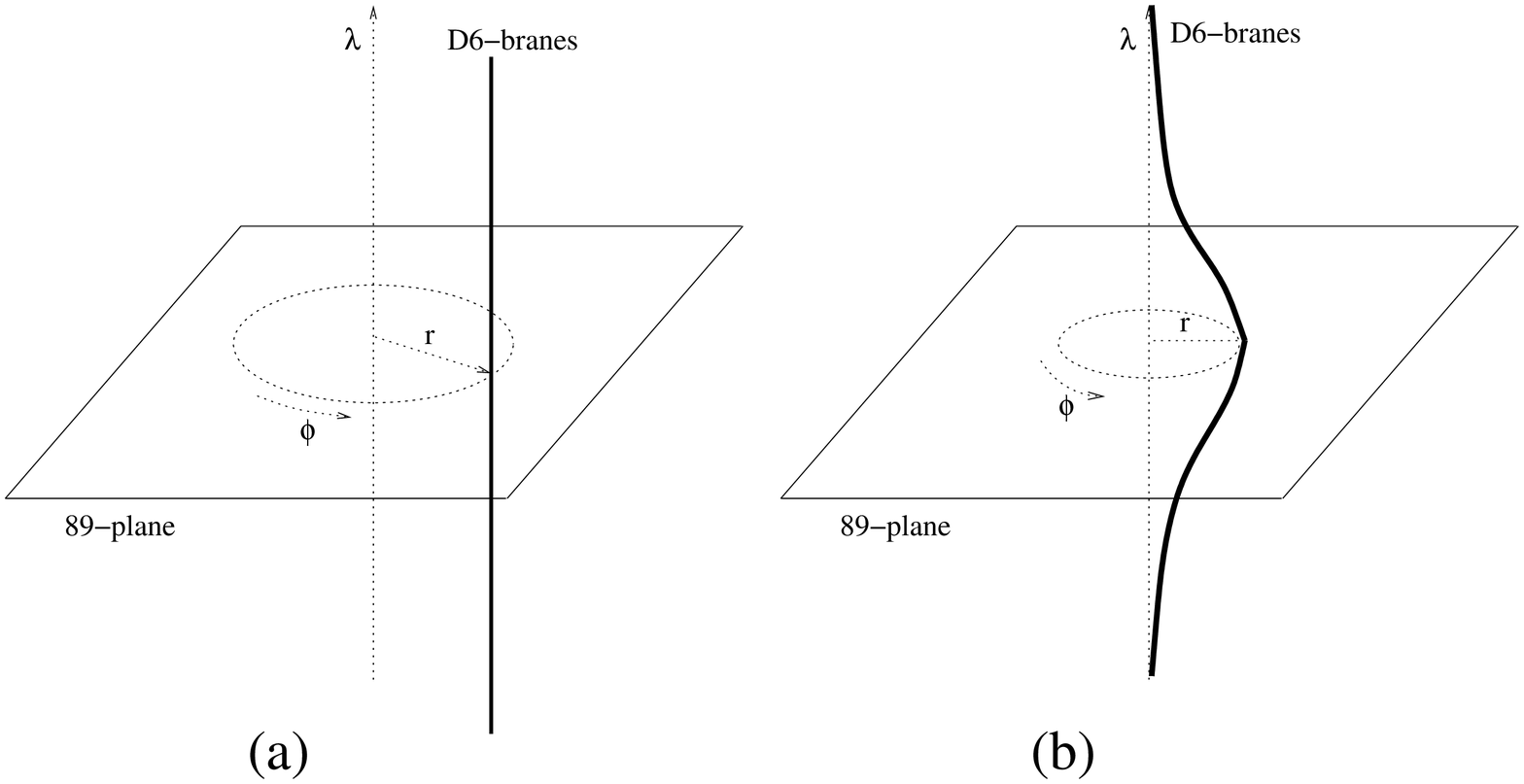, height=8cm}
\caption{(a) D6-brane embedding if $\ukk=0$, for some non-zero value
  of $\mq$. (b) D6-brane embedding for $\ukk \neq 0$ and $\mq=0$.}
\label{d6-embedding}}

If $\ukk \neq 0$ supersymmetry is broken and $r(\l)=\mbox{constant}$
is no longer a solution. The new solution is found as follows.
For large $\l$, the equation of motion linearizes, and its general
solution is
\be
r(\l) \simeq 2\pi\ls^{\,2}\,\mq + \fc{c}{\l} + O(\lambda^{-2}) \,.
\label{rt}
\ee
As explained in \cite{KMMW03}, the field $r(\l)$ is dual to the quark
bilinear operator $\bar{\psi} \psi$, so the constants $\mq$ and
$c$ are dual to the quark mass and the chiral condensate,
respectively. The requirement that the solution be regular everywhere
imposes a constraint between these two constants, that is, determines
$c=c(\mq)$. This is exactly as expected on field theory grounds, since the
chiral condensate should be dynamically determined once the quark mass
is specified.

The solution for massless quarks is depicted in figure
\ref{d6-embedding}(b).
We see that, although the D6-branes align asymptotically with
the $\l$-axis, they develop a `bump' in the 89-plane as
$\l \ra 0$, that is, $r(0) \neq 0$. The D6-brane embedding is
therefore not invariant under rotations in the 89-plane, and hence the
$\ua$ symmetry is {\it spontaneously} broken. The reason why this
breakng is spontaneous is that both the boundary condition,
$r(\infty)=0$, and the D6-brane equation of motion, are $\ua$-invariant,
yet the lowest-energy solution breaks the $\ua$ symmetry.
On gauge theory grounds, we expect this breaking to be caused by a
non-zero chiral condensate, $\cc \neq 0$. This is confirmed in the
gravity description by the fact that $c(\mq)$ approaches a non-zero
constant in the limit $\mq \ra 0$ \cite{KMMW03}.

The D6-brane embedding described above must be thought as the `vacuum
state' of the D6-branes in the D4-brane background. By studying
fluctuations around this embedding, the spectrum of (a certain class
of) scalar and pseudoscalar mesons was computed in \cite{KMMW03}.
In particular, for $\nf=1$, a massless, pseudoscalar meson was
found. This is the Goldstone boson expected from the spontaneous
breaking of $\ua$ symmetry, that is, the $\ep$. The corresponding
mode in the gravity description is the zero mode associated to
rotations of the D6-brane in the 89-plane, that is, it corresponds
to fluctuations of the D6-brane worldvolume field $\phi$.\footnote{
The odd-parity nature of these fluctuations is due to the fact that
a gauge-theory parity transformation acts on $X_8+ i X_9=r e^{i\phi}$
by complex conjugation. See \cite{KMMW03} for a detailed discussion.}

\section{The Anomaly-induced Potential and Glueball Mixings}
\label{anomaly}

In this section we discuss the general structure of $1/\nc$
corrections responsible for the generation of a potential that lifts
the $\ep$ meson. We first show that the introduction of D6-branes
corresponding to massless quarks allows the
$\theta$-dependence of the supergravity description to be shifted
away, as expected on field theory grounds. We then isolate the
relevant string diagrams and study their main properties in the
supergravity approximation.

\subsection{The anomaly relation in the ultraviolet regime}

At very high energies, the string model based on $\nc$ D4-branes and
$\nf$ D6-branes realizes the anomalous $\ua$ symmetry of QCD as an R-symmetry
on their common $\bbr{4}$ worldvolume. Since this symmetry is
anomalous, the
$\ua$ rotation of the D6-brane fields by an angle
$\alpha$, as specified in (\ref{rsim}), must be equivalent to a
shift of the effective
$\theta$-angle in (\ref{thec}) by
\be
\label{shift}
\int_{S^1} \cone \rightarrow \int_{S^1} \cone + \nf\, \alpha \,,
\ee
so that the dependence on the microscopic $\t$-angle can be eliminated
by a phase rotation  of the
$X_8+iX_9$ field, as argued
in the Introduction.

In the dual gravity description, the microscopic coupling (\ref{thec})
and the elementary quark fields $\psi_\mt{L,R}$ are not directly
visible, since the D4-branes are replaced by the background
\eqn{metric} and the effective action only contains colour-singlet degrees of
freedom. However, the fact that the dependence on the microscopic
$\t$-angle can be eliminated, as implied by the anomaly,
still follows from topological properties of the RR fluxes induced
by the D6-branes, as we now show.

In the gravity description, the microscopic $\t$-angle is defined by
the boundary condition \eqn{deft}. The key observation is that the
D6-branes' contribution to this integral has precisely the form
(\ref{shift}). To see this, we recall that, by definition, the
D6-branes are magnetic sources for the RR two-form, such that the
flux through any two-sphere that links the D6-branes is
\be
\label{links}
\int_{S^2} \ftwo = 2\pi\nf \;.
\ee
The D6-branes are localized in the $\tau$-direction, and, in the chiral
limit, they are also asymptotically localized at the origin of the
$89$-plane, \ie, $\lim_{\lambda\to \infty} r(\lambda)=0$. A two-sphere
surrounding the D6-branes in this region is shown in figure
\ref{flux}. Since $\tau$ is periodically identified, this two-sphere
can be continously deformed to a torus, $T^2$, parametrized by
$\tau$ and $\phi$ at fixed $r$ and (large) $\l$. Since $\ftwo$ is a
closed form, the captured flux is the same, \ie,
\be
\label{ff}
\int_{T^2} \ftwo = 2\pi\nf \;.
\ee
Since a translation in $\phi$ is an isometry of the background, it
follows that the flux through any strip defined by two angles
$\phi_1$ and $\phi_2$, as in the figure, must be proportional to the
area of the strip, that is,
\be
\int_{\mt{Strip}} \ftwo = \nf (\phi_2 - \phi_1) \;.
\ee
Note that this result relies crucially on the fact that all
integrals above are evaluated in the UV, \ie, in the limit
$\l \ra \infty$, as appropriate to the definition of the {\it
microscopic} $\t$-angle. In this limit the D6-branes lie at the
origin of the 89-plane and the integrals above are insensitive to
the deformation of the D6-branes in the region $\l \ra 0$.
\FIGURE{\epsfig{file=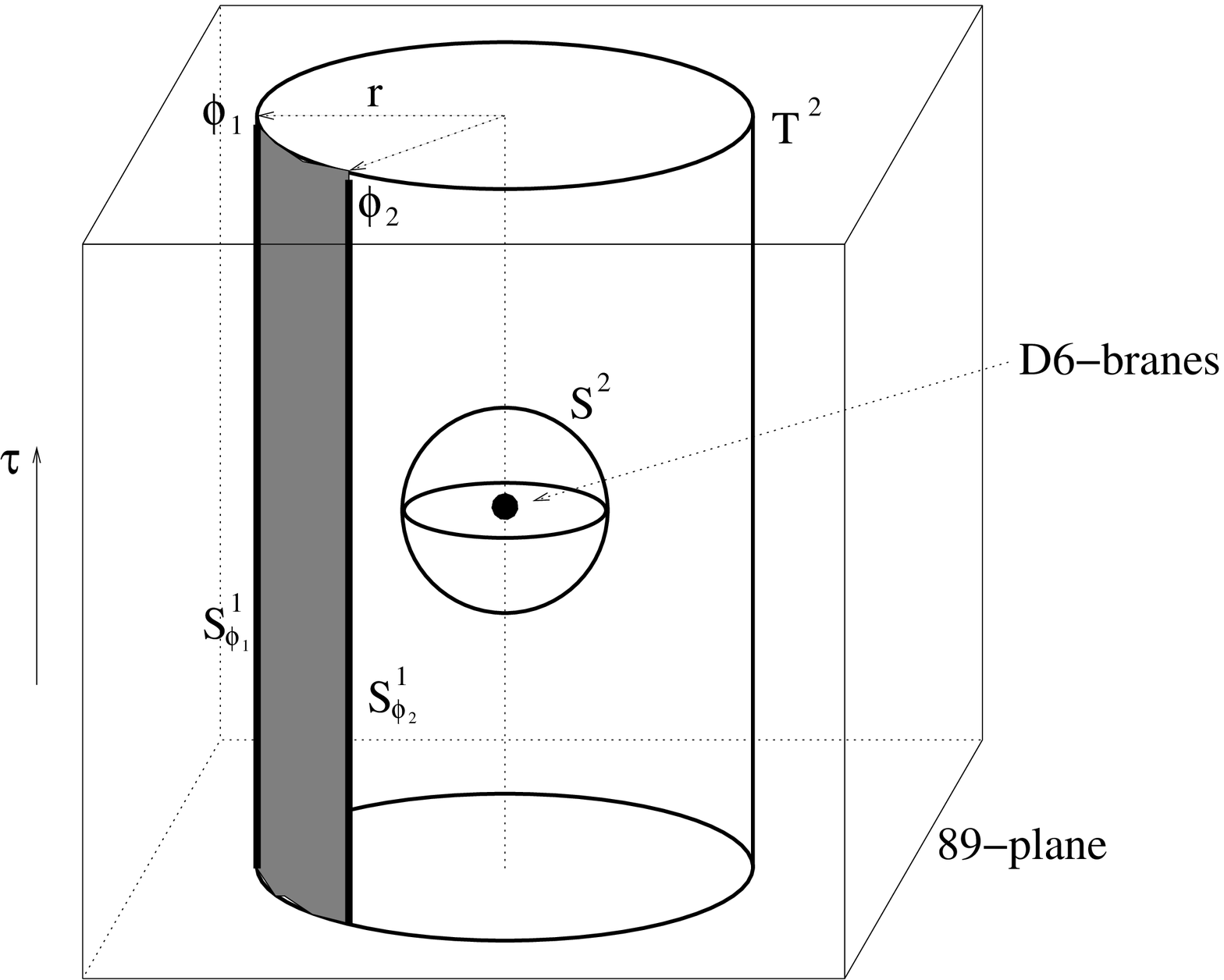, height=7cm}
\caption{Asymptotically, the D6-branes lie at the origin of the
89-plane and are localized in the $\tau$-direction.}
\label{flux}}
Finally, since locally we have $\ftwo=d\cone$, we can use Stokes'
theorem to write
\be
\int_{\mt{Strip}} \ftwo =
\int_{S^1_{\phi_\mt{2}}} \cone - \int_{S^1_{\phi_\mt{1}}} \cone \,,
\ee
where $S^1_{\phi_i}$ is parametrized by $\tau$ at $\phi=\phi_i$.
Combining these results we deduce that the Wilson line of $\cone$
at a given angle $\alpha$, as induced by the D6-branes, is
\be
\label{ind}
\int_{S^1_{\alpha}} \cone = \nf \, \alpha \,,
\ee
where we have set to zero a possible additive constant by choosing
the origin of the polar angle $\alpha$ appropriately.
If, in addition, there is a background value for this Wilson line
(an asymptotically flat connection defining the $\theta$-angle) then
the total value of the Wilson line is
\be
\label{tot}
\int_{S^1_\alpha} \cone = \theta + \nf \,\alpha \;.
\ee
Under a rotation by
angle $\alpha$ in the $89$-plane of the background, the `Dirac
sheet' singularity that is used to define $\cone$ (extending as a
string in the plane $(r,\phi)$ at $\phi=0$) rotates by minus this
same angle and shifts the theta angle according to (\ref{tot}).
 Since the position of this Dirac sheet is a gauge artefact, we see
explicitly how the microscopic $\t$-angle can be shifted away by a
$\ua$ transformation.

This supergravity argument proves that the physics is independent of
the microscopic $\t$-angle when the D6-branes are asymptotically
located at the origin of the $89$-plane, \ie, in the chiral limit.
Supersymmetry breaking at a scale $\mkk$ implies that a shift
$\delta \theta$ of the $\t$-angle by a change of the RR two-form $\ftwo$
costs energy $\chi_g \,\t \, \delta \theta$, to linear order in
$\delta \theta$. At the same time, chiral symmetry breaking implies that
a linear potential $\chi_g \,\t\, \nf \, \phi$ for the D6-brane
coordinate $\phi$ must be somehow generated, so that the complete
potential energy is only a function of the $\ua$-invariant
combination $\t + \nf \phi$. This is  checked in section
\ref{quanti} by an explicit computation.

 Since $\phi$ starts life in ten dimensions as a
gauge field, the mass term $\half \chi_g \,\nf^2 \phi^2$
 looks very much like a Green--Schwarz
correction to the field-strength of the $\cone$ axion field. It
would be interesting to confirm this by  finding a more geometrical
construction in ten-dimensional notation.

\subsection{String contributions to the potential}

The $\theta$-dependence computed by Witten in the pure-glue sector, plus
the above anomaly argument, constrain the leading potential of the $\ep$
field in the $k$-th branch to be
\be\label{potl}
V(\phi)^{(k)} = \half \chi_g\,(\theta + 2\pi k+  \nf\,\phi\,)^2 \,.
\ee
Mimicking the field theory arguments of
\cite{Witten79, Veneziano79} we can identify
the candidate string diagrams that generate the mass term by considering
string contributions to the two-point function of the {\it total} topological
susceptibility $\chi_{\rm total}$, which vanishes because of the anomalous
$\ua$ symmetry. In the string loop expansion, the pure-glue contribution
calculated in (\ref{tosu}) must be cancelled by contributions from meson
diagrams.
\FIGURE{\epsfig{file=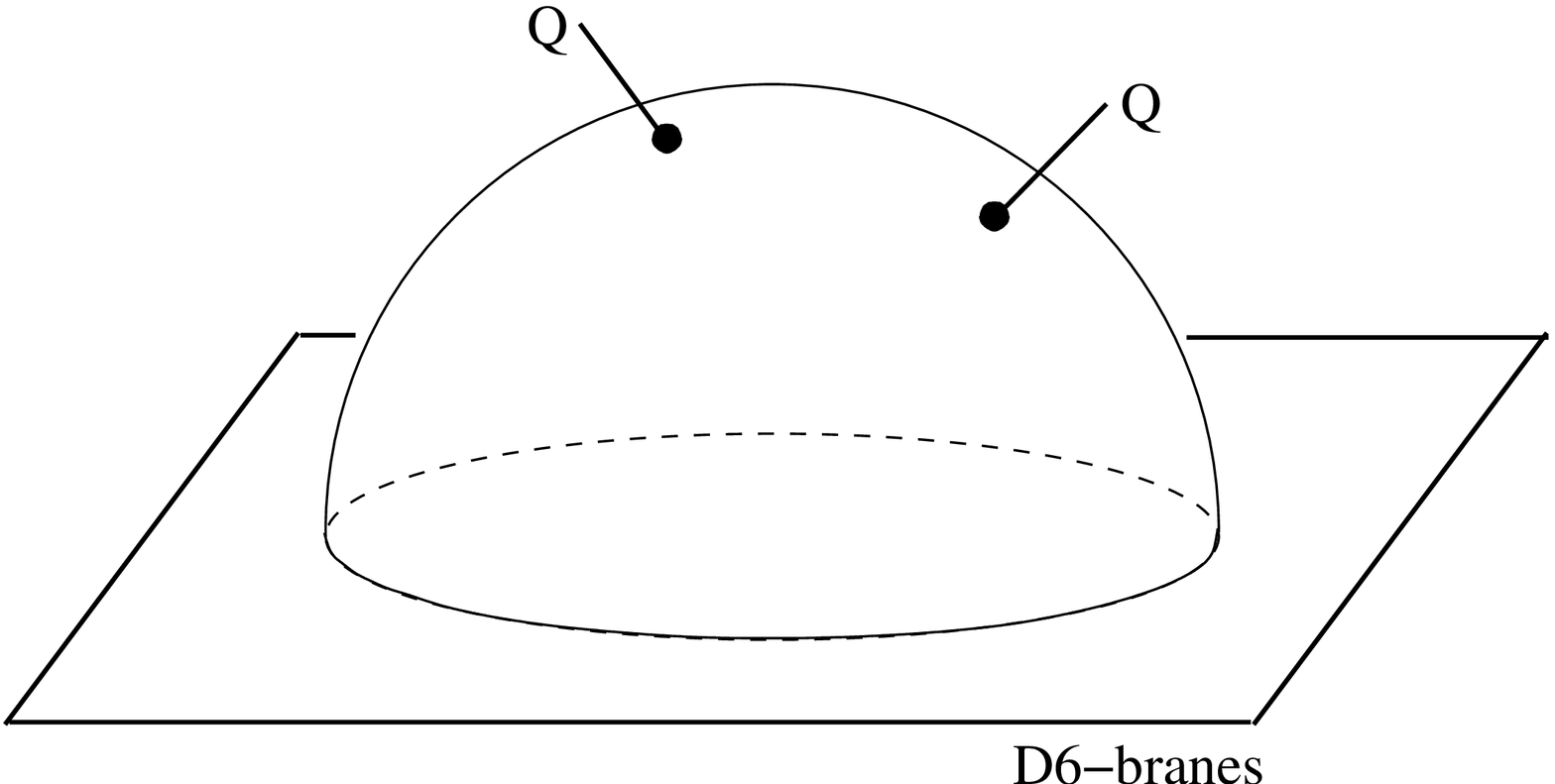, height=3.5cm, width=6.5cm}
\caption{The leading open-string correction to the two-point function
$U_1(p)$ of topological charge operators ${\cal Q} = \tr\,F\wedge F$.}
\label{disc}
}
The leading such diagram is depicted in figure \ref{disc}
and features a single open-string boundary attached to the D6-branes,
together with two closed-string vertex operators
dual to the anomaly operator $\calq = \tr\,F\wedge F$. This diagram is the
string counterpart of the OZI-suppressed quark annhilation diagrams
considered in \cite{RGG75, Witten79, Veneziano79}.

A spectral decomposition of this diagram yields
\be
\label{spec}
U_1 (p) = {\nf \over \nc} \sum_n {|C_n|^2 \over p^2 + m_n^2}\;,
\ee
where $C_n = O(1)$ in the large-$\nc$ limit and the meson spectrum
$m_n$, calculated from the fluctuations of the D6-brane, is also of
$O(1)$, except for the lowest excitation, the $\ep$, which is massless.
The contribution to the topological susceptibility arises from the
formal $p \rightarrow 0$ limit, which of course is infrared-divergent
because of the massless $\ep$ meson.
\FIGURE{\epsfig{file=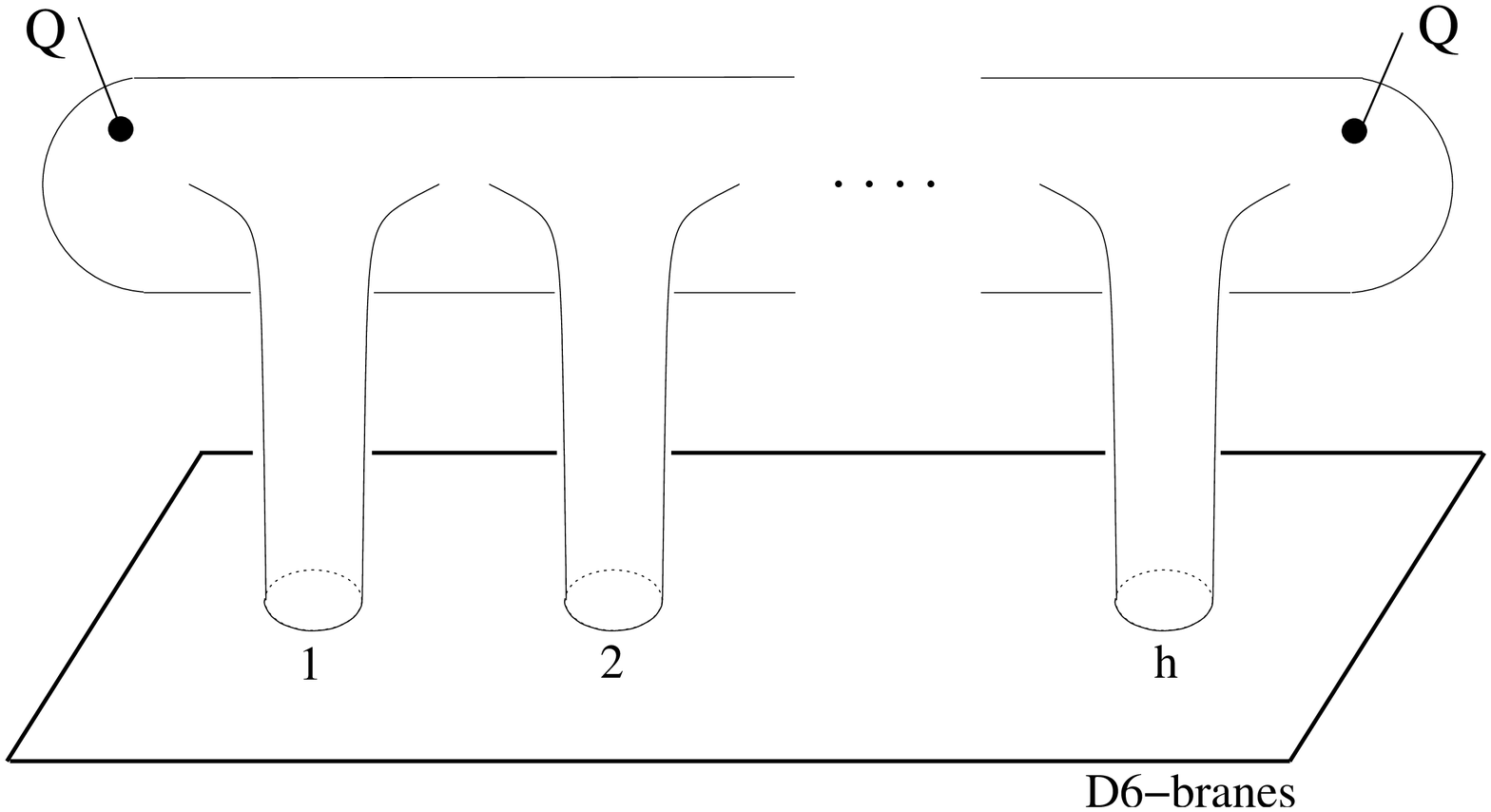, height=3.5cm, width=6.5cm}
\caption{Diagram $U_h (p)$ with $h$ open-string boundaries.}
\label{sum}
}

A standard procedure to resolve this infrared divergence is to resum
a chain of highly divergent diagrams, $U_h (p)$, of the form
depicted in figure \ref{sum}, where the index $h$ stands for the
number of open-string boundaries. Isolating the massless meson in
$h$ intermediate propagators, we see that $U_h (k)$ diverges in the
infrared as $(\nf /  \nc p^2 )^h$. Summing up the geometric series
of such terms induces a 1PI self-energy contribution, of order $\nf
/\nc$, given by the cylinder diagram in figure \ref{cylinder}(a).
The same diagram with the other possible inequivalent insertion of
the $\eta'$ field (contributing of course self-energy corrections of
the same order) is depicted in figure \ref{cylinder}(b).

The closed and open string interpretations\footnote{By this we mean
  those obtained by cutting the
  diagrams in such a way that the intermediate states are closed or open
  strings, respectively. Of course, in both cases the external states
  are an open string state, namely, the $\eta'$.} of these diagrams is
given in figures \ref{closed} and \ref{open}, respectively. Note
that the indices carried by the double lines are not $SU(\nc)$ colour
indices (there are only $SU(\nc)$ singlets in the gravity description)
but flavour indices of $SU(\nf)$,\footnote{This is a global symmetry
  of the boundary field theory, and a gauge symmetry on the
  worldvolume of the D6-branes in the dual gravity description.}
under which the pion fields transform in the adjoint representation
but the $\eta'$ is inert.

As shown by its open string representation, the diagram in figure
\ref{cylinder}(b) is equivalent to a standard one-loop correction in
the effective meson theory, and this contributions is common to
singlet and non-singlet mesons. In contrast, the diagram in figure
\ref{cylinder}(a) will only couple
to the flavour singlet mesons and so distinguishes the behaviour
of the $\eta'$ meson from the rest of the `Goldstone' modes.
This would suggest that, at a quantitative level, this diagram gives
the most important contribution to the mass of the $\eta'$.

In order to contribute to the $\ep$ mass, either of these self-energy
corrections must shift the zero-momentum pole of the large-$\nc$
meson propagator. Unfortunately, direct computation of the full string
diagrams is not possible in the background in question, since we are
restricted to the supergravity approximation. It is then interesting to
separate the part of figure \ref{cylinder}(a) corresponding to the
exchange of supergravity modes from a stringy `contact term' coming
from the infinite tower of closed string modes and possible
contributions at the boundary of worldsheet moduli space.
The contribution of a {\it finite} number of low-lying glueball modes
with mass $M_n$ shifts the $\eta'$ pole mass-squared by
\be\label{mixshift}
\delta m_\ep^2 = - \sum_n {g_n (0)^2 \over M_n^2}
\;,
\ee
where $g_n (0)$ stands for the zero-momentum limit of the
glueball-$\ep$ coupling, which must be non-vanishing for this
contribution to be non-trivial.  The shift (\ref{mixshift}) has
the `wrong' sign though, so the stringy contact term (the high-energy
part of the full string diagram) must be positive and all-important
at the quantitative level.

We will elaborate further on these issues in the last section.

\FIGURE{\epsfig{file=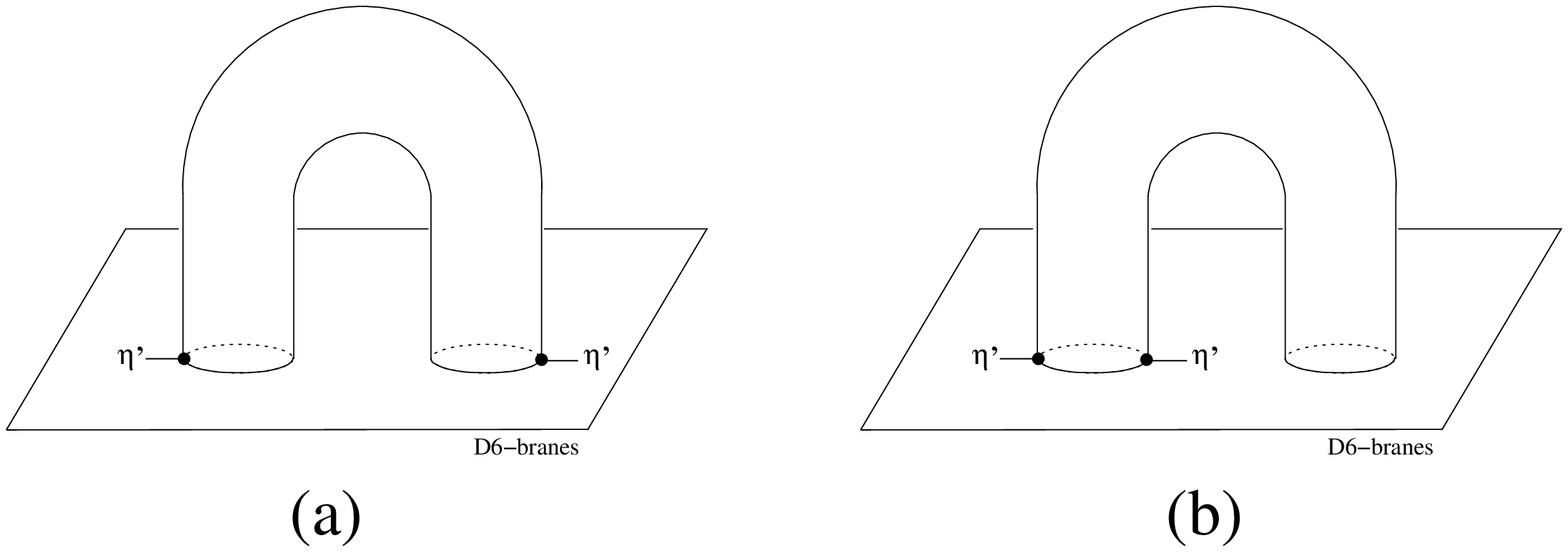, width=12cm}
\caption{Basic cylinder diagram of order $\nf/\nc$, with the two possible
inequivalent insertions of the $\eta'$ field.}
\label{cylinder}}
\FIGURE{\epsfig{file=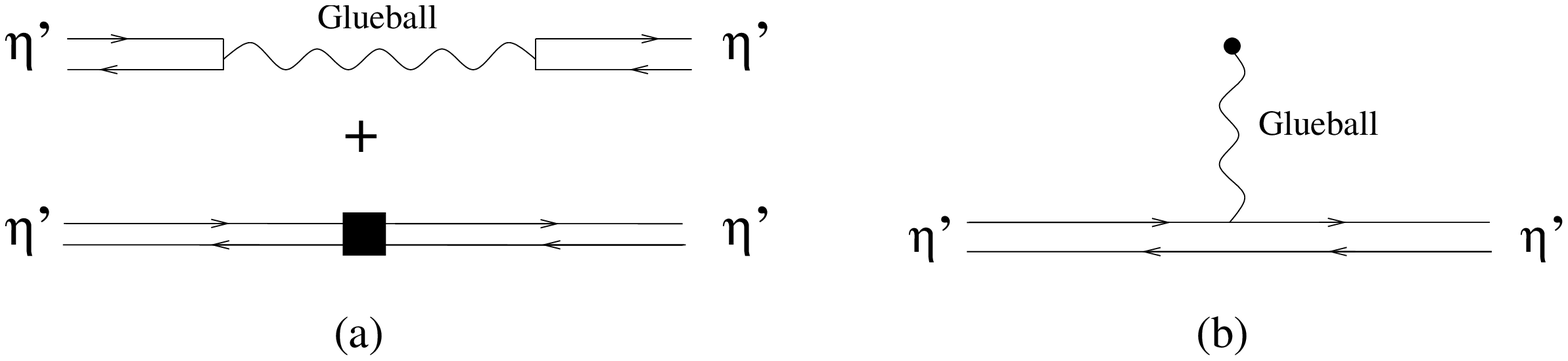, width=12cm}
\caption{Closed string interpretation of the cylinder diagrams of
  figure \ref{cylinder}. The representation
  (a) exhibits the fact that diagram \ref{cylinder}(a) contributes
  low-lying glueball mixing at tree level (supergravity fields)
  plus a high-energy contact term coming from the infinite tower of closed
  string
  states. The $\eta'$-glueball coupling in (a) is of order $\sqrt{\nf/\nc}$.
  The strength of the glueball tadpole in (b) is of order $\nf$,
  whereas the cubic $\eta'$$\eta'$-glueball vertex is of order $1/\nc$
  --- see Appendix A.}
\label{closed}}
\FIGURE{\epsfig{file=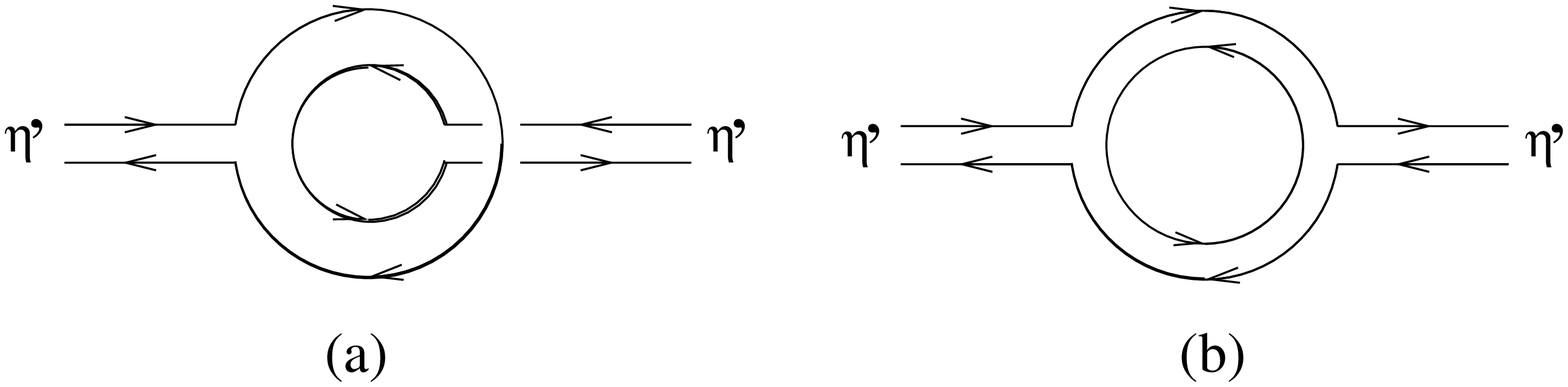, width=12cm} \caption{Open
string interpretation of the cylinder diagrams of figure
\ref{cylinder}. (b) is the standard one-loop meson self-energy; each
of the two vertices is of order $1/\sqrt{\nc}$, and the internal
loop yields a factor of $\nf$. The internal meson propagators in (a)
are both twisted. The factor of $\nf$ now comes from the fact that
the flavour of the incoming lines need not be the same as that of
the outgoing lines.}
\label{open}}

\subsection{Meson-glueball mixing}

In this subsection we show that $g_n (0) \neq 0$ by analysing the
glueball-$\ep$ mixing at the supergravity level.
Quite generally, any closed-string field $\Phi_c$ that
is sourced by the D6-branes and has non-trivial wave-function with respect
to the $\phi$ angle is subject to mixing with the $\ep$ meson.
Expanding $\Phi_c$ in Fourier modes one has
\be
\label{expp}
\Phi_c(\phi) = \sum_n {\cal G}_n \,e^{-in\phi}
\;,\ee
where the normalizable modes ${\cal G}_n$, when pulled back to the
$\bbr{4}$ factor in the D6 world-volume, represent  glueballs of
$\ua$ charge $n$. If $\Phi_c$ enters linearly the world-volume theory
on the D6-branes, equation (\ref{expp}) gives the required non-derivative
couplings to the $\ep$ meson. The prototypical example is the dilaton
term in the Born-Infeld action:
\be
\label{bi}
{1 \over (2\pi)^6 \,\ls^{\,\,7} } \sum_{j=1}^{\nf}
 \int_{\Sigma_7} e^{-\Phi} \,\sqrt{-\det G_\mt{ind} (\partial \phi_j)}
\;,
\ee
where $\Sigma_7$ is the D6-branes worldvolume,
\be
\label{ex}
\sqrt{-\det G_\mt{ind} \,(\partial \phi_j)} =
1 + O\left[(\partial \phi_j)^2 \right]
\ee
and the corresponding pullbacks are understood to each of the coinciding
$\nf$ branes. Selecting the non-derivative term in the expansion of
the square root we have couplings of the form (\ref{expp}).
In chiral-lagrangian notation, assembling the collective coordinates
of the D6-branes in a Goldstone-boson diagonal matrix
$\Sigma = {\rm diag}\,(e^{i\phi_j})$, we have terms of the form
\be\label{cl}
\nc \sum_n \int_{\bR^4} {\cal G}_n \,\tr\,\Sigma^n + {\rm h.c.}
\ee
These glueballs, being charged with respect to the $\ua$ group, are
`Kaluza--Klein artefacts', not present in real QCD. In fact, the couplings
(\ref{cl}) respect the $\ua$ symmetry and cannot induce a potential that
breaks it upon integrating out the glueballs. For example, at tree level
we generate terms proportional to $\tr\,\Sigma^n \cdot \tr\,\Sigma^{-n}$,
because the glueball propagator couples ${\cal G}_n$ and ${\cal G}_{-n}$
($n$ being a Kaluza--Klein momentum). The global phase $e^{i\phi}$ drops
from these expressions and we see that such couplings do not generate
a potential for the $\ep$ particle. This is just as well, since such
contributions seem completely independent of the $\t$-dependence, as dictated
by the Witten--Veneziano formula.

In fact, the candidate glueballs are selected by the general arguments
in the Introduction. First, we expect the required couplings to show the
characteristic multivaluedness of $\t$-dependence at large $\nc$, \ie, we
expect the coupling to be a function of $-i\,\log\,{\rm det}\,\Sigma \sim
\nf\,\phi$, precisely linear in the angular coordinate, so that the
angular periodicity of the effective action would require an explicit sum over
different branches.  The supposed relation to $\t$-dependence suggests
that we investigate the glueballs in the RR sector of the closed-string theory.

Natural candidates are the normalizable modes of the RR potential
$\cone$, or, equivalently, of its Hodge dual $\csev$, since these give
rise to {\it pseudo}-scalar glueballs. In the absence of D6-brane
sources, it is truly equivalent to work with $\cone$ or $\csev$. The
D6-branes, however, couple minimally to $\csev$ through the
Wess--Zumino term
\be
\label{wzt}
S_\mt{WZ} = {\nf \over (2\pi\ls)^6} \int_{\Sigma_7} \csev \,.
\ee
In terms of $\csev$, this coupling is both local and can be defined
for off-shell values of the RR seven-form potential. For on-shell configurations
this coupling can be reexpressed in terms of $\cone$ at the expense of
introducing non-locality. However, since we wish to exhibit
$\eta'$-glueball couplings at zero momentum, which are necessarily
off-shell, we must work with $\csev$.

Now we wish to demonstrate that \eqn{wzt} contains a linear coupling to
the $\phi$ field when reduced to the $\bbr{4}$ factor of the space-time.
Towards this end, let us consider the following ansatz for fluctuations:
$\csev = (\phi+\phi_0) W_{\it 7}$,
where $\phi_0$ is a constant and $W_{\it 7}$ is the $\phi$-independent
seven-form
\bea
W_{\it 7} = &&- G(x) \, h(U) \, r \l^2 \, (r d\l - \l dr) \w
d\Omega_{\it 2} \w dV_{\it 4} \nn
&& + \tilde{h}(U) \, r\l^2 \, d\l \w dr \w d\Omega_{\it 2} \w
i_{N(x)} dV_{\it 4} \,.
\label{W}
\eea
Here $d\Omega_{\it 2}$ and $dV_{\it 4}$ are the volume forms on the $S^2$ wrapped
by the D6-branes and on the $\bbr{4}$ factor, respectively.
$G(x)$ is a pseudoscalar field and $N^\mu(x)$ is a vector field with

\be
i_{N(x)} dV_{\it 4} =
\fc{1}{3!} N^\mu (x) \, \e_{\mu\nu\a\b} \, dx^\nu \w dx^\a \w dx^\b \,.
\ee
Finally, $h(U)$ and $\tilde{h}(U)$ are radial profiles to be determined.
Note that $\csev$, not being gauge-invariant, is
allowed to be multivalued in $\phi$,\footnote{An average over the action
  $\phi_0 \ra \phi_0 +2\pi$ can restore the angular character of
  $\phi$ that is lost in the expression for $\csev$, and is the
  counterpart of the average over large-$\nc$ branches of
  $\t$-dependence.}
but its gauge-invariant field strength, $F_{\it 8}=d\csev$, must be
single-valued. This restriction
forces $W_{\it 7}$ to be closed, which in turn implies
\be
G(x) \, \left[ 5h(U) + \rho \, \fc{dU}{d\rho} \, h'(U) \right] +
\pa_\mu N^\mu (x) \, \tilde{h}(U)=0 \,,
\ee
and therefore
\be
\tilde{h} = 5h + \rho \, \fc{dU}{d\rho} \, h' \sac \pa_\mu N^\mu = - G \,.
\label{cons}
\ee
We see that closure of $W_{\it 7}$ relates the two radial functions,
as well as the scalar and the vector. Under these conditions
$F_{\it 8} = d\phi \w W_{\it 7}$, which is, of course, single-valued.

We may regard the second equation above as a constraint on $N$,
and $G$ as a yet totally unconstrained pseudoscalar glueball field.
In fact, $N$ is not an independent field
on-shell, but is completely determined by $G$. Indeed, the equations
of motion for $G$ and $N$ come from $d * F_{\it 8}=0$, or
equivalently the Bianchi identity
$d F_{\it 2} =0$. A straightforward calculation yields
\be
F_{\it 2} = * F_{\it 8} = G(x) \, \tilde{H}(U) \, \rho
\left(\fc{dU}{d\rho} \right)^{-1} \, d\tau \w dU -
H(U) \, N_\mu \, d\tau \w dx^\mu \,,
\ee
where
\bea
\tilde{H}(U) &=& - \left( \fc{U}{R} \right)^{-9/4} \, K(\rho)^{-3/2} \,
f(U)^{1/2} \, h(U) \,, \nn
H(U) &=& \left( \fc{U}{R} \right)^{-3/4} \, K(\rho)^{-5/2} \,
f(U)^{1/2} \, \tilde{h}(U) \,,
\eea
and we have made use of the fact that
\be
dU = \fc{dU}{d\rho} \left( \fc{\l}{\rho} \, d\l +
\fc{r}{\rho} \, dr \right) \,.
\ee
Closure of $F_{\it 2}$ then implies
\be
H' = M^2 \, \tilde{H} \, \rho \, \left(\fc{dU}{d\rho} \right)^{-1}
\sac N_\mu = -\fc{1}{M^2} \, \pa_\mu G
\label{all}
\ee
for some constant $M^2$. As anticipated, the second equation above
determines $N$ in terms of $G$. Combined with the constraints
\eqn{cons}, it imposes  the on-shell condition for $G$:
\be
\pa_\mu \pa^\mu \, G = M^2 \, G \,.
\ee
Further combining the first constraint in \eqn{cons} with
the first equation in \eqn{all} yields a second-order ODE for
the radial profile. This equation provides an eigenvalue problem
that determines the pseudoscalar glueball mass spectrum, $M_n^2$,
as well as the corresponding normalizable radial profiles,
$h_n, \tilde{h}_n$. Once these profiles are known, the
non-derivative $\phi$--$G_n$ couplings arise from the Wess--Zumino term
by pulling back $\csev$ onto the D6-branes worldvolume and reducing
the result along the $S^2$ and along the radial direction down to
four-dmensions. The coordinate $\phi$ is pulled back into a field
$\phi_0+\phi(x)$ that depends {\it only} on the $\bbr{4}$ coordinates.
The form $W_{\it 7}$ is pulled-back on the ground state of the
D6-branes embedding, $\tau=0, r=r(\l)$, since we are only interested
in the couplings of the $\eta'$ and not the rest of the mesons.
The non-derivative couplings originate from the first summand in
$W_{\it 7}$, and take the form (setting $\phi_0=0$)
\be
\label{mixc}
S_\mt{WZ} \ra {\nf f_\pi \over 2\sqrt{\nc}} \,
g_n \,\int_{{\bR}^4} \phi(x) \, G(x)
=\sqrt{\nf \over \nc} \, g_n \int_{{\bR}^4} \eta'(x) \,
G(x) \,,
\ee
where
\be
\label{intf}
{f_\pi \over 2\sqrt{\nc} } \;g_n = {{\rm vol}(S^2)
 \over (2\pi\ls)^6} \int_0^\infty d\lambda\;
h_n (U(\lambda)) \, r^2 \l \,\left( r-\lambda {\dot r} \right) \,,
\ee
and we recall that $\phi$ and $\eta'$ are related as in \eqn{cann}.
In principle, these couplings can be evaluated numerically, given the
embedding $r(\l)$ and the eigenmode profiles $h_n(U)$. However, the
fact that they are in general non-vanishing is already
an important result, for it confirms that the
cylinder diagram in figure \ref{cylinder}(a) is capable of generating a
potential for the $\ep$ with the right properties.

\section{A Quantitative Check  to Order $1/\sqrt{\nc}$}
\label{quanti}

We have argued in the previous sections that certain quantum corrections to
the supergravity model with probe D6-branes \cite{KMMW03} generate a
potential for the $\ep$ meson of the form
\be
\label{thep}
V(\ep\,) = \half\,\chi_g\,\left(\theta +
{2\sqrt{\nf} \over f_\pi}\,\ep\,\right)^2
\ee
to leading order in the $1/\nc$ expansion. In this expression, we have considered
the $k=0$ branch of the vacuum energy and we have fixed the additive normalization
of the $\ep$ field so that $V(\ep =0)$ equals the pure-glue vacuum energy
derived in equation \eqn{wr}. With these conventions, taking into account that
$f_\pi = O(\sqrt{\nc})$, we can expand the square and separate the pure-glue
term of $O(1)$, the Witten--Veneziano mass term of $O(\nf/\nc)$, and a
cross term of $O(\sqrt{\nf /\nc})$ which acts as a tadpole upon expanding
the potential around the wrong vacuum, $\ep=0$.
In this section we present a calculation of this linear term by two independent
methods, one based on a closed-string calculation plus the anomaly argument,
and the other based on a direct open-closed string coupling.

In terms of the $\phi$ field, the `tadpole' term can be identified as
\be\label{tadpole}
{\rm tadpole} ={\cal T}= \chi_g\,\t\,\nf\, \phi\;,
\ee
and we may evaluate it in two independent ways. First, we can use the
explicit supergravity calculation \eqn{tosu} of the pure-glue
topological susceptibility and introduce the $\phi$-dependence
via the anomaly argument $\t \rightarrow \t + \nf\,\phi$. We find
\be
\label{tuno}
{\cal T} = {C \, \theta \,\nf \over 3\cdot 2^4 \cdot \pi^5\cdot\ls^{\,\,6}}
\, \phi \,,
\ee
where we remind the reader that $C= 3\, \ukk^3 /\d \tau$ from
(\ref{wittf}). We emphasize that this calculation only uses the
closed-string sector, plus the microscopic anomaly argument.

On the other hand, we may read the linear term directly from the
Wess--Zumino action (\ref{wzt}) for the particular seven-form
$\csev$ that is induced by the $\theta$-angle background \eqn{wittf}.
Setting $k=0$ in this equation, a straight-forward calculation shows
that the dual seven-form potential is given (locally) by
$\csev = \phi \, \osev$, where
\be
\label{ome}
\osev = {C \, \theta \over U^4} \,B(U) \,(r\,d\lambda - \lambda\,dr)
\wedge d\Omega_{\it 2} \wedge dV_{\it 4} \,,
\ee
where
\be
B(U) = \fc{1}{\rho} \, \fc{dU}{d\rho} \,
\left( \fc{U}{R} \right)^{9/4} \, \fc{K(\rho)^{3/2}}{f(\rho)^{1/2}}
\l^2 r \,,
\ee
and we have set to zero the additive normalization of $\phi$, as well as
the discrete $2\pi$-shift implementing the large-$\nc$ branches of vacua.
Calculating the pull-back of $\csev$ on the D6 world-volume,
as in the previous section, we obtain
\be
\label{calta}
{\nf \over (2\pi\ls)^6} \int_{\Sigma_7} \phi \,\osev =
\int_{\bR^4} {\cal T} \,,
\ee
where
\be
\label{tadd}
{\cal T} = C\, \t \, \nf \, \phi \,
{{\rm vol}(S^2) \over (2\pi\ls)^6} \int_0^\infty d\lambda
{H(U(\lambda)) \over U(\lambda)^4} \,\left(r-\lambda {\dot r}\right)
\ee
or, using $\rho^2 (\lambda) = r^2 (\lambda) + \lambda^2$,
\be
\label{ct}
{\cal T}= {C \, \theta \, \nf \over 2^4 \cdot \pi^5\cdot\ls^{\,\,6}} \, \phi \,
\int_0^\infty d\lambda {\lambda^2 r \over \rho^5} (r-\lambda {\dot r})
\,.
\ee
Agreement with \eqn{tuno} requires the last integral to equal
$1/3$. Remarkably, this is so, for the integral can be transformed into
\be
\int_0^\infty d\l {\l^2 \over \rho^4} \,(\rho-\l {\dot \rho}) =
\fc{1}{3} \int_0^\infty d\l \, \fc{d}{d\l} \left( \fc{\l^3}{\rho^3}\right)
= \fc{1}{3} \,,
\ee
where we have used in the last step the fact that $r(\lambda)$ remains
bounded as $\lambda \rightarrow \infty$ in the D6-brane embedding.

A more geometrical version of this calculation  can be given as
follows. We have argued that, locally in the $\phi$-direction,
$\csev = \phi \,\osev$. The tadpole comes just from the integration of
$\osev$ over $\Sigma_7$, the equilibrium worldvolume of the D6-branes
at $\phi=0$. Now, let $\Sigma_8$ be the hypersurface that results from
rotating the worldvolume around the angle $\phi$. Since
$\partial_\phi\, \osev =0$ for the $\theta$-induced form \eqn{ome},
we can write
\be
\int_{\Sigma_7} \osev = {1\over 2\pi} \int_{\Sigma_8} d\phi \wedge \osev
= {1\over 2\pi} \int_{\Sigma_8} \feig \,.
\ee
In addition, $\Sigma_8$ is topologically equivalent to the $S^4$
at fixed $U$ coordinate, times the spacetime $\bbr{4}$ factor. Since
$\feig$ is a closed form, we can use Stokes' theorem to write
\be
\int_{\Sigma_7} \osev= {1\over 2\pi} \int_{\bR^4 \times S^4} \feig \,.
\ee
The latter integral is trivially evaluated by computing $\feig$
directly in the original coordinate system, in which it takes the
simple form
\be
\feig = *\,\ftwo = C\,\theta\,d\Omega_{\it 4} \w dV_{\it 4} \,.
\ee
It follows that the prediction for the tadpole is
\be
\label{tadp}
{\cal T} =  \phi\cdot {\nf \over (2\pi\ls)^6} \cdot
{C\theta \over 2\pi} \cdot {\rm vol}(S^4) =
{C\, \theta\,\nf \over 3\cdot 2^4 \cdot \pi^5\cdot\ls^{\,\,6}} \, \phi\,,
\ee
again in perfect numerical agreement with (\ref{tuno}). We regard this
check as highly non-trivial, since the kinetic RR term only knows
about closed strings (glueballs) and the Wess--Zumino term specifies the
direct coupling to the open strings (mesons). The exact agreement for
the tadpole is an indication that the basic physical picture is right.

\section{Concluding Remarks}

In \cite{KMMW03} a string dual of large-$\nc$ QCD with $\nf$
flavours, based on $\nf$ D6-brane probes in a fixed supergravity
background, was studied in detail. It was found that the string
description captures some of the low-energy physics expected on
field theory grounds. In particular, for $\nf=1$, it exhibits
spontaneous chiral symmetry breaking of an $\ua$ symmetry, and the
mesonic spectrum contains a pseudoscalar that is exactly massless at
infinite $\nc$. This is the analog of the $\ep$ meson of large-$\nc$
QCD, and is dual to the zero-mode associated to the motion of the
D6-brane.

As discussed in the Introduction, the importance of the
Witten-Veneziano formula (\ref{wv}) is in producing a qualitative
understanding of the mass splitting of the $\eta'$ meson from the
other light mesons, namely the pseudo-Goldstone modes associated
with the spontaneous breaking of the chiral flavour symmetry. With
multiple flavours, that is, multiple D6-branes, our AdS-like model
produces $\nf^2$ massless pseudoscalars \cite{KMMW03}. However,
only the diagonal mode corresponding to the collective
center-of-mass motion of all the D6-branes is obviously a
Goldstone mode. In Appendix B, we argue that, when the analysis is
taken beyond tree-level, the other $\nf^2-1$ modes acquire masses
of order $(\lambda \nf / \nc)^{1/2} \mkk$, where $\lambda =
\gym^2 \nc$ is the 't Hooft coupling.
As further shown in Appendix B, this precisely matches the
mass of the $\eta'$ at the level of their parametric dependences.
We interpret the masses of the off-diagonal modes as arising from
closed string interactions between the individual D6-branes and
so, to leading order, they are generated by the same string
diagrams as illustrated in figure
\ref{cylinder}(b) -- recall only the diagonal mode couples in
figure \ref{cylinder}(a). This result points to a qualitative
distinction between the physics of our model and QCD: even in the
limit of vanishing quark masses, our entire multiplet of $\nf^2$
light mesons acquires a mass squared of the same order as the
$\ep$, while in QCD only the $\ep$ becomes massive. The additional
mass terms in the present case are natural as the off-diagonal
scalars are not Goldstone modes --- in fact, it is the
masslessness of these modes in the large-$\nc$ limit that was
surprising \cite{KMMW03}. However, it would still be interesting
to refine the estimates made in Appendix B, to see if there is any
dramatic difference in the numerical values of the off-diagonal
and the $\ep$ masses for our model.

On general grounds, the identification by Witten \cite{Witten98b}
of an $O(1)$ contribution to the pure-glue topological susceptibility,
plus a microscopic anomaly argument, implies the generation of a potential
for the $\eta'$ with a mass term of $O(\nf/\nc)$, along the lines
of the Witten--Veneziano argument. We have argued that the
cylinder diagram of figure \ref{cylinder}(a) is the relevant stringy
correction responsible for the generation of the $\ep$ mass, in a
string analog of the Isgur-de R\'ujula-Georgi-Glashow mechanism
\cite{RGG75}. We have shown that, in the supergravity approximation,
this diagram induces non-derivative mixings between the glueballs and
the $\ep$ that shift the zero-momentum pole of the $\ep$. However,
this shift by itself would make the $\ep$ tachyonic, and hence we
argued that there is an important contact term
coming from the stringy completion of the
glueball-exchange diagrams in figure \ref{cylinder}(a). This
discussion is natural when one thinks of the worldsheet cylinder
as being long in comparison to its circumference. Coming from the
opposite end of the worldsheet moduli space ({\it i.e.,} a short
cylinder with a large circumference), this diagram has a natural
interpretation, depicted in figure \ref{open}(a), in terms of
an open string, and hence meson, loop,
where the internal meson propagators are now both twisted. From
this point of view, which naturally figures as the UV completion of
the glueball sum, these loop contributions do not have a definite
sign and so certainly allow for the necessary shift with a
positive sign. Of course, string duality tells us that that the
sums over all tree-level glueball exchanges and over one-loop
meson graphs are the same and should not be computed separately.

Similarly, the open string or meson loop of figure \ref{open}(b)
also has an interpretation as a
closed string coupling with two $\eta'$ mesons and being absorbed
by the D6-brane, as displayed in figure \ref{closed}(b).
It might be emphasized here that both diagrams in figure \ref{closed}
play an important role in the $\eta'$ physics. This is
particularly evident from the discussion at the beginning of
section 3.3, where we argue that the Neveu-Schwarz glueballs will {\it not}
generate a potential for the $\eta'$. To see how the cancellation
presented there occurs order by order in $\phi$, one would make a
Taylor expansion of the exponentials in equation (\ref{expp}). At
order $\phi^2$, this reveals that the vanishing mass arises
precisely as a cancellation between glueball exchange as in figure
\ref{closed}{a} and tadpole contributions of figure \ref{closed}(b).

More generally, the anomaly argument implies that the dependence of
the theory on the $\t$-angle occurs only through the combination
$\t + 2 \sqrt{\nf} \, \ep / f_\pi$, so that the microscopic $\t$-dependence
can be eliminated by a $\ua$ transformation.
 We have verified this statement in
the ultraviolet regime by analizing the RR flux sourced by the
D6-branes.

We have strenghened the physical picture by performing a
quantitiative check of the $\eta'$-potential at order
$1/\sqrt{\nc}$. We have computed this potential in two independent
ways. One method employs only the closed string sector, that is, the
pure-glue sector, together with the anomaly argument. The second
method involves the open string sector, that is, the mesonic sector.
We regard the perfect agreement between the two results as a
non-trivial check that the right physics is captured.

The `master substitution' $\t \ra \t + 2\sqrt{\nf} \,\ep/f_\pi$
applied to the pure-glue effective Lagrangian generates all
soft-$\ep$ amplitudes. In the supergravity formalism, this implies
precise correlations between the effective couplings of the closed
string sector, and those of the closed string sector to the open
string sector. The simplest of these correlations was checked in
section \ref{quanti}, but it would be interesting to investigate the
more complicated ones, even at a qualitative level.

Since the pure-glue $\theta$-dependence comes from the energy of RR
fluxes, the stringy mechanism is akin to a Green--Schwarz modification
of the RR field strengths.\footnote{See \cite{adi} for a study of
  this interpretation in a slightly different model.}
It would be interesting to sharpen the anomaly argument in the
supergravity regime by identifying a ten dimensional anomaly polynomial
that yields the substitution $\t \ra \t + 2 \sqrt{\nf} \, \ep / f_\pi$
as a standard Green--Schwarz modification of the $\ftwo$ field strength,
after appropriate reduction on the D6-branes worldvolume. In this case,
the basic stringy calculation of the ultraviolet contact terms could
be performed locally in the flat limit of the
ten-dimensionsional string theory.

\acknowledgments

We would like to thank  Margarita Garc\'{\i}a P\'erez,
C\'esar G\'omez and Sean Hartnoll for helpful
discussions, and especially Adi Armoni
for many discussions on the subject of this paper and \cite{adi}.
The work of J.L.F.B. was partially supported by MCyT
and FEDER under grant BFM2002-03881 and the European RTN network
HPRN-CT-2002-00325. The work of C.H. was partially supported by
European Commission (HPRN-CT-200-00148) and CICYT (Spain) and
by the MECD (Spain) through a FPU grant. Research at the Perimeter
Institute is supported in part by funds from NSERC of Canada.
RCM is further supported by an NSERC Discovery grant.

\appendix

\section{Large-$\nc$ Scalings from Supergravity and DBI}

In the presence of $\nf$ D6-branes, the low-energy dynamics on the
gravity side is described by the action
$S=S_\mt{Sugra}+S_\mt{D6}$. Let $\Phi$ and $\phi$ denote collectively
fluctuations of the supergravity and the D6-branes' worldvolume fields,
respectively. Schematically, the action for these fields takes then
the form
\be
S = \nc^2 \, \int_{M_\mt{10}} (\pa \Phi)^2 + \sum_{\ell\geq 2} \Phi^\ell
+ \nf \nc \int_{\Sigma_7} (\pa \phi)^2 +
\sum_{m \geq 0, n \geq 1} \Phi^m \, \phi^n \,.
\ee
The coefficients in front of each integral arise from the scalings
with $g_s$ of Newton's constant, $G_N \sim g_s^2$, and the fact that
$g_s \sim 1/ \nc$ in the 't Hooft limit. Further the
D6-branes' tension scales as $T \sim \nf /g_s$ --- note that we are
assuming that $\phi$ is a $U(1)$ field in the $U(\nf)$ gauge group
on the D6-branes, as is the $\eta'$.

In terms of canonically normalized fields, defined through
$\Phi \ra \nc \Phi$ and $\phi \ra \sqrt{\nf \nc} \phi$, we find
\be
S = \int_{M_\mt{10}} (\pa \Phi)^2 + \sum_{\ell\geq 2} \nc^{2-\ell} \Phi^\ell
+ \int_{\Sigma_7} (\pa \phi)^2 + \sum_{m \geq 0, n \geq 1}
\nf^{1-\fc{n}{2}} \nc^{1-m-\fc{n}{2}} \Phi^m \, \phi^n \,.
\ee
The strength of the different couplings can now be directly read
off. For example, a glueball tadpole (\ie, a closed string one-point
function) is of order $\nf$ while the $\eta'$-glueball coupling is of order
$\sqrt{\nf/\nc}$. These results are used in section 3.2.

\section{Pseudoscalar Masses Revisited}

In this paper, we have argued that the $\ep$ meson, in the model
of \cite{KMMW03}, acquires a mass consistent with the
Witten-Veneziano formula:
\be \label{wv2} m_\ep^2 = {4 \nf \over f_\pi^2} \,\chi_g \,. \ee
We now wish to evaluate this mass in terms of the microscopic
parameters of the field theory. In the following, we will only
determine the parametric dependence but drop numerical factors.
First, equation (\ref{tosu}) gives the topological susceptibility
as
\be \label{sus2} \chi_g\sim\l^3\mkk^4 \ee
where $\l=\gym^2\nc$ is the 't Hooft coupling. From (4.18) of
\cite{KMMW03}, one deduces that, up to numerical factors, the pion
decay constant is given by
\be \label{fpi} f_\pi^2\sim T_\mt{D6} \ukk^3/\mkk^2 \sim
\nc\l^2\mkk^2\,. \ee
The second expression above was determined using (\ref{inverse})
and $T_\mt{D6}\sim1/g_s \ls^7$. Combining these results, we find
\be \label{wv3} m_\ep^2 \sim {\nf\over\nc}\,\l\,\mkk^2\,. \ee

With a collection of $\nf$ D6-branes in the holographic model of
\cite{KMMW03}, the dual theory contains $\nf$ quark flavors. If
all of the quark masses vanish, it was shown to leading order in
the large-$\nc$ expansion that the spectrum contains $\nf^2$
massless pseudoscalar mesons. Of these, only the $\ep$ meson,
which corresponds to the collective motion of all of the D6-branes
together, appears to be a true Goldstone mode. Again the latter
only applies in the large-$\nc$ limit, as we have argued here that
a mass appears through $1/\nc$ effects. Similarly it was argued in
\cite{KMMW03} that the remaining $\nf^2-1$ pseudoscalars will
acquire masses at this order. These mesons can be interpreted as
modes which separate the individual D6-branes and so the masses
can be understood as arising from closed string interactions
between the separated branes. We will now give a concrete (albeit
crude) estimate of the masses which are produced in this way. In
the following, we will be working with the background D4-brane
metric in the form given in eq.~(\ref{isometric}).

\FIGURE{\epsfig{file=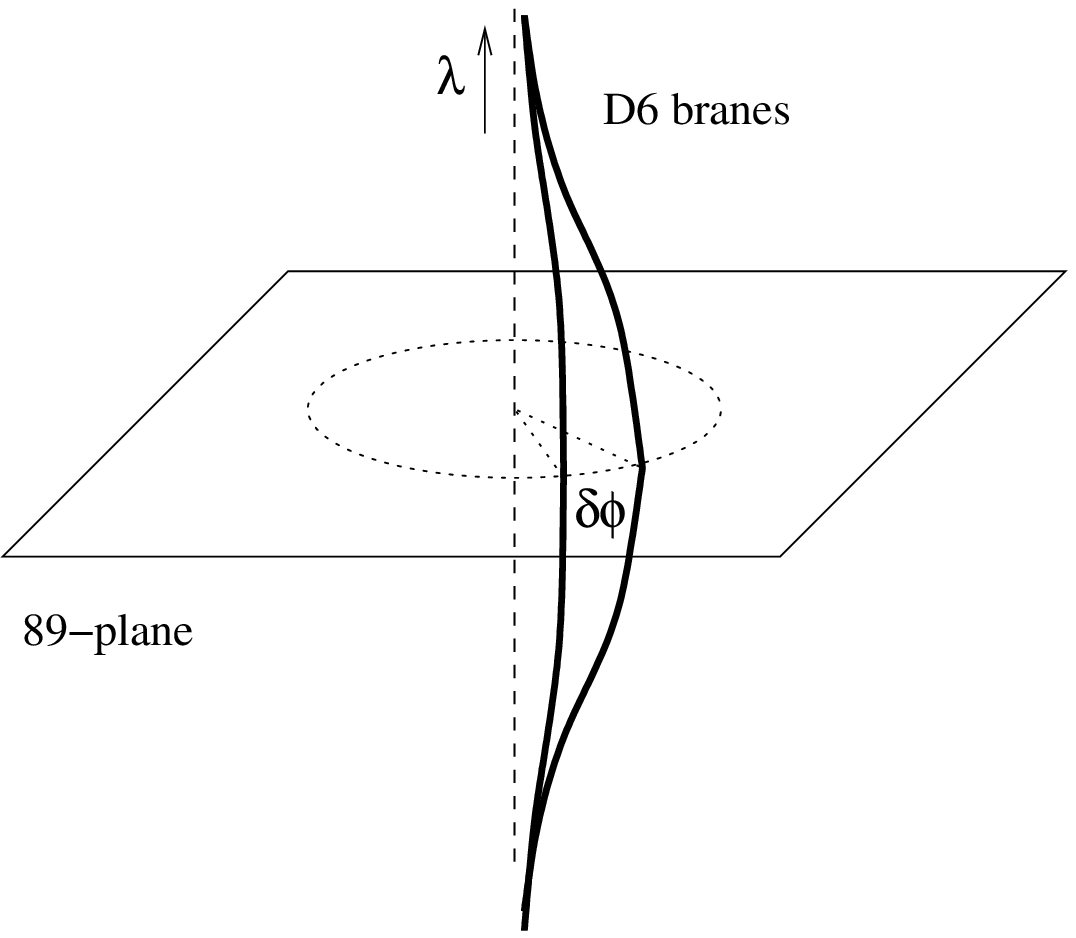, width=12cm} \caption{D6-brane
embedding for $\ukk \neq 0$ and $\mq=0$, but separating two groups
of $\nf/2$ D6-branes by a small angle $\delta\phi$ in the
89-plane.} \label{figure8}}

Imagine we have embedded $\nf$ D6-branes with $\mq=0$ but they
have been divided into two groups of $\nf/2$ branes separated by a
small angle $\delta\phi$ in the 89-plane, as depicted in figure
\ref{figure8}. We will assume this depicts a canonical mode and so
its mass is typical of that for all of the off-diagonal modes. We
will also assume that the interaction can be approximated by
integrating the `Newtonian' potential between volume elements on
the separated (sets of) branes. Now as all of the branes have the
123-space in common, it is simplest to consider integrating over
slices which extend through this subspace. These three-dimensional
volume elements are then only infinitesimal in their extent in the
$z^i$ subspace where the embedding is nontrivial, and will
interact with a $1/\rho^4$ potential. We express the interaction
energy as the energy density per unit volume (in the 123-space),
which will roughly take the form:
\be \delta{\cal E} \simeq \nf^2\,G_{\it 10}\, h(\Omega_1,\Omega_2)
{(T_{\mt{D6}}\,\delta^3A)_1\,(T_{\mt{D6}}\,\delta^3A)_2\over
|\vec{z}_1-\vec{z}_2|^4}\ ,\label{pot3}\ee
where we have introduced $h(\Omega_1,\Omega_2)$ to indicate that
the strength of the interaction depends on the details of the
relative orientation of the two area elements.\footnote{Implicitly
we will assert below that this interaction strength is positive.
This intuition comes from considering D-branes in flat space where
static parallel branes are supersymmetric. Hence a relative
rotation between elements of the brane can only increase the
energy.}

Now if we examine eq.~(2.17) of \cite{KMMW03} which determines the
embedding profile of the D6-branes in the background geometry, we
see that $\ukk$ is the only scale which enters this equation and
so this is the scale of the interesting deformation in figure 8.
Hence, we expect that the appropriate integrations yield
\be \left\langle h(\Omega_1,\Omega_2)
{(\delta^3A)_1\,(\delta^3A)_2\over |\vec{z}_1-\vec{z}_2|^4}
\right\rangle\sim \ukk^2\,\delta\phi^2\ , \ee
up to an overall purely numerical factor. That is, $\ukk$ and
$\delta\phi$ are the only relevant scale and
angle\footnote{Symmetry would rule out the appearance of a single
power of $\delta\phi$ and hence the leading contribution must be
$\delta\phi^2$.}, respectively, in the geometry of the D6-brane
configuration. However, we must also account for the fact that all
proper distances in the $z^i$ subspace are contracted by a factor
of $K^{1/2}$, as seen in eq.~(\ref{isometric}). Hence the energy
density acquires an additional factor of
\be
K={R^{3/2}U^{1/2}\over\rho^2}\sim\left({R\over\ukk}\right)^{3/2}\
, \ee
where in the last expression we have again used the approximation
that $\ukk$ is the only scale relevant for the embedding. Next, we
note that the final expression may be simplified using
\be G_{\it 10}\, T_\mt{D6}^2\simeq
g_s^2\ls^8\left(1/g_s\ls^7\right)^2=1/\ls^6\ .\ee
Hence our approximation for the total interaction energy density
becomes
\bea {\cal E}&\sim& \nf^2
{R^{3/2}\ukk^{1/2}\over\ls^6}\,\delta\phi^2\nonumber\\
&\sim&\nf^2{\lambda\over\ls^4}\,\delta\phi^2 \eea
where we simplified the second line with (\ref{inverse}).

Now one might find the appearance of $\ls$ in this expression
disturbing, however, one must realize that this is not a field
theory quantity rather it is a proper energy density, ${\cal
E}_{proper}$. To convert our expression to the energy density in
the dual field theory, we must include additional metric factors
for the $x^\mu$ directions relating proper bulk space distances to
simple coordinate distances, which are relevant for the field
theory, \ie, from eq.~(\ref{isometric}), $\Delta x_{proper}\sim
\left(\ukk/R\right)^{3/4}\Delta x_{coord}$ again making the
approximation that the only scale relevant for the embedding is
$\ukk$. Hence we have
\be {\cal E}_{field}\sim \left({\ukk\over R}\right)^3{\cal
E}_{proper}\sim \nf^2\lambda^3\mkk^4\,\delta\phi^2 \ee
where the latter expression is simplified using (\ref{inverse}).
Finally to identify the mass, we must normalize the coefficient of
$\delta\phi^2$ above by that in front of the corresponding kinetic
term. Following the discussion of \cite{KMMW03}, it is
straightforward to see that the kinetic term for this off-diagonal
mode has an overall factor of $\nf f_\pi^2$. Hence, using
eq.~(\ref{fpi}), our final expression for the mass is
\be \label{wv4} m_{pseudo}^2 \sim {\nf\over\nc}\,\l\,\mkk^2\,. \ee
Hence the parametric dependence of these off-diagonal pseudoscalar
masses precisely matches that of the $\ep$ in eq.~(\ref{wv3}). It
may be interesting to perform a more detailed analysis to
determine the numerical coefficients in these two mass formulae.

\end{document}